\documentclass[fleqn,10pt]{wlscirep}
\usepackage[utf8]{inputenc}
\usepackage[T1]{fontenc}
\usepackage{subcaption}
\DeclareMathOperator*{\argmin}{arg\,min}
\usepackage{tabularx}

\title{A Study on The Effectiveness of Lock-down Measures to Control The Spread of COVID-19}

\author[1,a]{Subhas Kumar Ghosh}
\author[2]{Sachchit Ghosh}
\author[3]{Sai Shanmukha Narumanchi}

\affil[1]{Commonwealth Bank of Australia, Sydney, New South Wales, 2000, Australia}
\affil[2]{The University of Sydney, Camperdown, NSW 2006, Australia}
\affil[3]{Department of Computer Science, Southern Illinois University, Carbondale, IL 62901, USA}

\affil[a]{subhas.ghosh@cba.com.au}


\begin{abstract}
	The ongoing pandemic of coronavirus disease 2019-2020 (COVID-19) is caused by Severe Acute Respiratory Syndrome Coronavirus 2 (SARS-CoV-2). This pathogenic virus is able to spread asymptotically during its incubation stage through a vulnerable population. Given the state of healthcare, policymakers were urged to contain the spread of infection, minimize stress on the health systems and ensure public safety. Most effective tool that was at their disposal was to close non-essential business and issue a stay home order. In this paper we consider techniques to measure the effectiveness of stringency measures adopted by governments across the world. Analyzing effectiveness of control measures like lock-down allows us to understand whether the decisions made were optimal and resulted in a reduction of burden on the healthcare system. In specific we consider using a synthetic control to construct alternative scenarios and understand what would have been the effect on health if less stringent measures were adopted. We present analysis for The State of New York, United States, Italy and The Indian capital city Delhi and show how lock-down measures has helped and what the counterfactual scenarios would have been in comparison to the current state of affairs. We show that in The State of New York the number of deaths could have been 6 times higher, and in Italy, the number of deaths could have been 3 times higher by 26th of June, 2020.
\end{abstract}

\begin{document}
\flushbottom
\maketitle
%
%
\thispagestyle{empty}
\section*{Introduction}
	\label{SEC1}
	In December 2019, an outbreak occurred in Wuhan, China involving a zoonotic coronavirus, similar to the SARS coronavirus and MERS coronavirus  \cite{taaa021}. The virus has been named Severe Acute Respiratory Syndrome Coronavirus 2 (SARS-CoV-2), and the disease caused by the virus has been named the coronavirus disease 2019 (COVID-19). Since then the ongoing pandemic has infected more than 9 million people and has caused more than 467 thousand deaths worldwide. Since the initial outbreak, several different studies have tried to estimate the number of infections \cite{GN2020} that stem from a single infected patient in order to predict the potential for transmission of the COVID-19 virus. In most cases, it was seen that $R_0 > 1$, implying exponential growth through infection of a vulnerable population. Original estimates placed mortality rates for individuals at high risk at  4.46 \% with those suffering from cardiovascular or kidney disease having even greater susceptibility \cite{BPH2020}. The SARS-CoV-2 virus has no available treatment as the pathways for proliferation and pathogenesis are still unclear \cite{RIS2020}.  Current treatments are based on those effective on strains of the previous SARS coronavirus and MERS coronavirus. The SARS-CoV-2  virus is able to replicate rapidly during the asymptomatic phase and affect the lungs and respiratory tract, resulting in pneumonia, hypoxia, and acute respiratory distress \cite{PSL2020}. Infected patients are directly dependent on external ventilation in most cases. 
	
	With the increasing pressure on the health systems due to reliance on intensive care units or non-invasive ventilation, health strategies were required to be implemented. The concern was to ensure the number of infected patients does not exceed the health system’s ability to cope with it. It also focused on increasing the capacities of available health systems at the time. Under the conditions at the time, with a highly pathogenic SARS-CoV-2 that is able to spread asymptotically during its incubation stage through a vulnerable population, policymakers were urged to contain the spread of the infection, and minimize stress on the health systems and ensure public safety. This was done by issuing orders for widespread lock-down and implementing social distancing measures. All non-essential businesses and services were shut down until further notice. 
	
	Taking measures to reduce stress on the health sector and diminishing the number of infected patients is important to end the pandemic, and understanding the effectiveness of a lock-down enables the distinction of good safety measures from bad ones. Analyzing effectiveness of control measures like lock-down allows us to understand whether the decisions made were optimal and resulted in a reduction of burden on the healthcare system, and broke chain of transmission, preventing its spread and reducing the reproductive rate of the virus. Any optimal policy considers a trade off between the benefit associated with lock-down and cost of reduced aggregate output. Aggregate output decreases as a function of the stringency of the policy, commitment from the government to maintain the level of stringency and adherence of general population. Aggregate output decreases through lower supply of labor, lower consumption and hence through lower investment, which results from investors’ expectation of a lower marginal product of capital. On the other hand benefit associated with lock-down can be seen through the number of lives potentially saved and in curbing the pandemic early so that economic activity can be restarted early. 
	
	Our objective in this work is to understand the benefits obtained from stringency measures adopted by governments across the world in terms of its health benefits. In the remaining of this section we describe our contribution and related  works. Subsequently in Section \ref{SEC2} we provide a brief overview and mathematical underpinning of the tools that we use and describe our data driven methodology and in Section \ref{SEC4} we present our results in three different geographic setup. Finally, in Section \ref{SEC5} we present some concluding remarks.
	
	\subsection*{Our contribution}
	In this work we consider stringency measures adopted by governments across the world and provide a counterfactual assessment of the benefit from those measures in terms of health benefits. In order to estimate the counterfactual metric (say number of deaths), we use a geographic location as a treatment unit (say Italy) and a set of other geographic locations as donor group (say Brazil and United States). We take a data driven approach to construct a synthetic control \cite{ap08746, JMLR18, AMSS19} using pre-intervention period data (say from early February, 2020 to March 9, 2020) of the donor units and their linear combination, such that the squared error between the estimated synthetic control  and the treatment unit is minimized by the choice of weight parameters in this pre-intervention time period. Now synthetic control can be extrapolated to estimate the metric. We use Multi-dimensional Robust Synthetic Control (m-RSC) as a tool as described in \cite{AMSS19}. 
	
	However, there are few difficulties in applying the tool as it is.  Firstly, different governments adopted different levels of stringency measures and there were different levels of compliance, and commitment. Secondly, there were no `pure` donor groups as stringency measures were nearly ubiquitous. So we have used various secondary sources of data to score the level of stringency measures , and level of compliance. This allowed us to determine donor groups relative to a choice of treatment unit.
	
	Finally, we present our results in three different geographic setup - namely in the State of New York, Italy and Indian city of Delhi, and analyze them. 
	
	\subsection*{Related Works}
	In \cite{Koh2020}, authors have shown that the reproductive rate of the SARS-CoV-2 had significantly decreased after government intervention. They show that the spread of disease was confined if measures were brought into effect early. In \cite{KP2020}, authors use the differential timing of the introduction of stringency measures and changes in Google searches for unemployment claims to establish a framework to estimate how each stringency measure contributes to unemployment. Authors show that early intervention efforts in the form of non-essential business closure have contributed to less than 8.5 percent of unemployment claims.
	
	Another facet to measure the success of the lock-down is to observe its effects on the health systems. Late intervention in the case of Italy led to the flooding of hospitals and ICUs due to exponential spread. However, the national lock-down was effective in reducing the proliferation and decreased the stress on the national health system as observed by authors in \cite{Supino2020}. In their paper \cite{IVEGA2020},  authors extend the SIR model to include auxiliary state variables in the form of hospital capacity, contact with an infected person, etc. They use a system dynamics model of the outbreak to simulate various lock-down scenarios with recommendations for optimal strategy. In our work we consider the possible outcome if such strategies were not adopted,  and present counterfactual scenarios.
	
	International travel has also been impacted as a result of efforts to reduce the spread of the coronavirus disease. On the basis of reported cases, models built by \cite{Chinazzi395} show a significant decrease in the number of infections compared to predictions if no travel bans were adopted as an option. Their modeling results indicate that travel restriction must be combined with a transmission within the community to curb the spread.

\section*{Results}
\label{SEC4}
In this section we present three examples of the application of m-RSC to derive the counterfactual estimates of possible number of deaths under the changed conditions like delaying or starting the stringency measures at earlier date. We consider three different units of treatments, namely: The State of New York, Italy and The Indian capital city Delhi. In some sense these places have also been termed as regional epicenters  of the epidemic.

\subsection*{State of New York}
New York has the highest number of confirmed cases in the United States. First case in New York was reported on 1st March, 2020 and New York went into a stricter lock-down on March 22nd, 2020. We estimate counterfactual considering this as date of intervention. We select among other states from US as donor group using the methods described above. This includes New Jersey, California, Illinois, and Florida among other states. By counterfactual estimate, the number of deaths in New York could have been 6 times higher, and number of confirmed cases could have been 5 times higher.

\begin{figure*}
	\centering
	\begin{subfigure}[b]{\textwidth}
		\centering
		\includegraphics[width=0.9\linewidth,height=0.2\textheight]{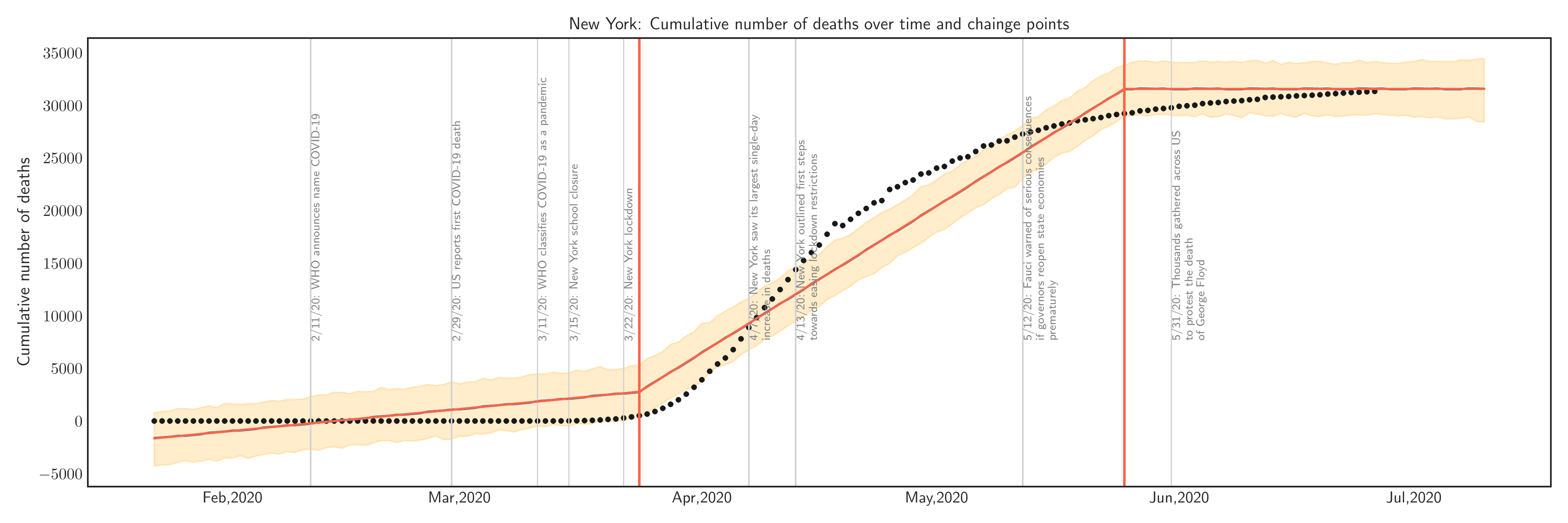}
	\end{subfigure}
	
	\begin{subfigure}[b]{\textwidth}
		\centering
		\includegraphics[width=0.9\linewidth,height=0.2\textheight]{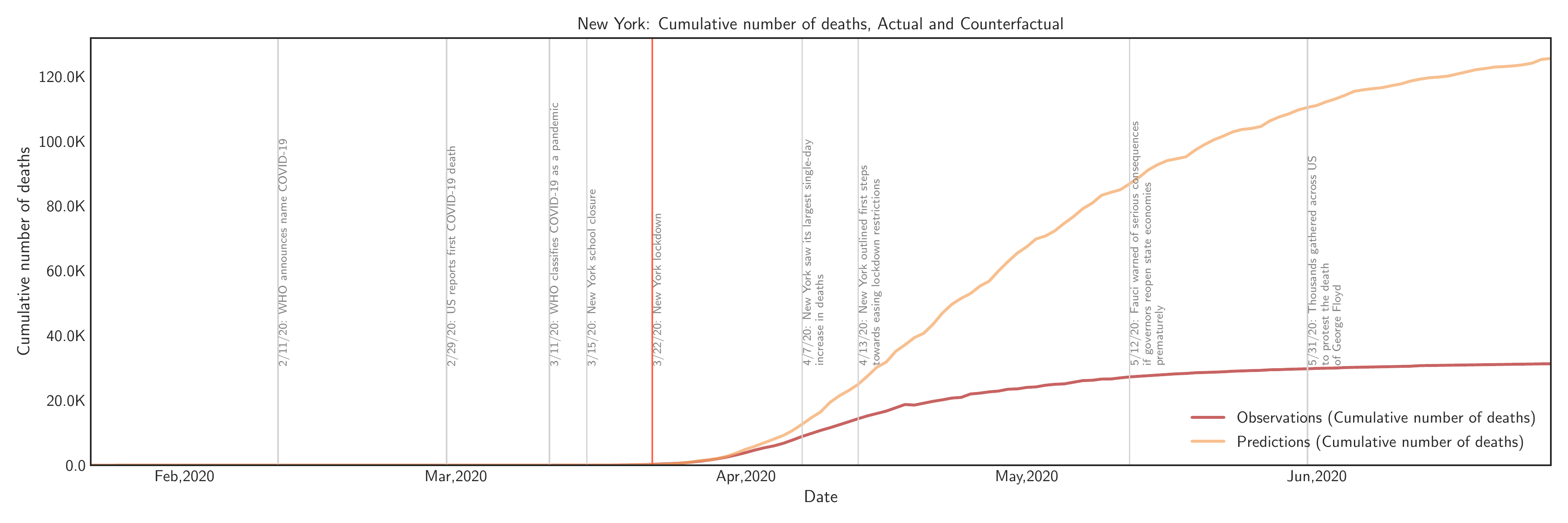}
	\end{subfigure}
	
	\begin{subfigure}[b]{\textwidth}
		\centering
		\includegraphics[width=0.9\linewidth,height=0.2\textheight]{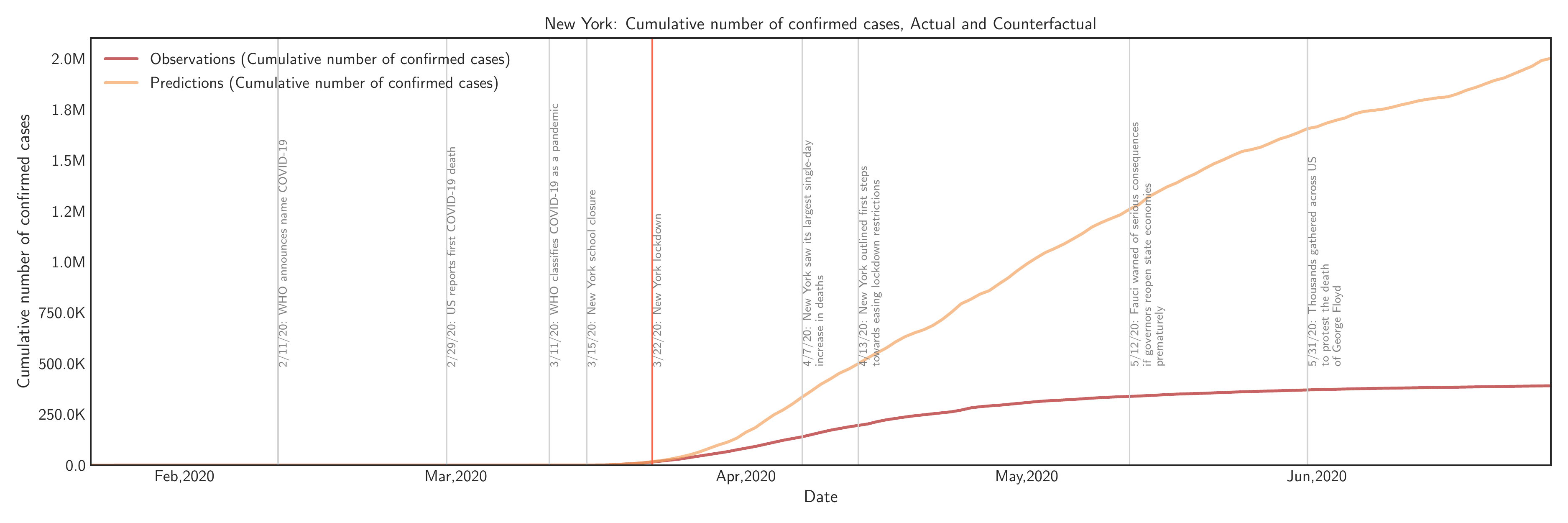}
	\end{subfigure}
	
	\caption[New York]{New York State: (a) Actual number of deaths with linear trend line and points where trend changed (b) Actual number of deaths with counterfactual prediction estimated by m-RSC with actual lock-down date as intervention point (c) Actual number of confirmed cases with counterfactual prediction estimated by m-RSC.}
	\label{fig5} 
\end{figure*}

\subsection*{Italy}
Italy was put under lock-down between 8th March, 2020 - 4th May, 2020. We considered most European countries to model the donor group and selected based on criteria defined above. Based on our simulation, we observe that with lock-down measures  has been largely successful in Italy. Without such measures, the number of confirmed case could have been 8 times higher and number of deaths could have been 3 times higher by 26th of June, 2020.

\begin{figure*}
	\centering
	\begin{subfigure}[b]{\textwidth}
		\centering
		\includegraphics[width=0.9\linewidth,height=0.2\textheight]{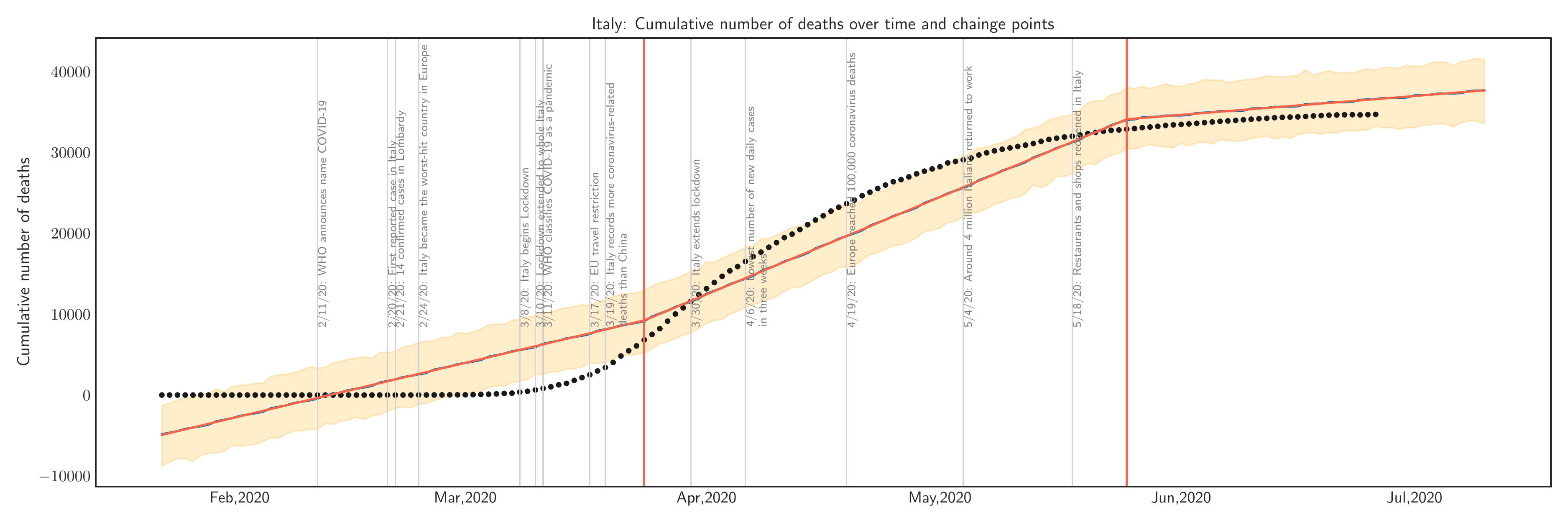}
	\end{subfigure}
	
	\begin{subfigure}[b]{\textwidth}
		\centering
		\includegraphics[width=0.9\linewidth,height=0.2\textheight]{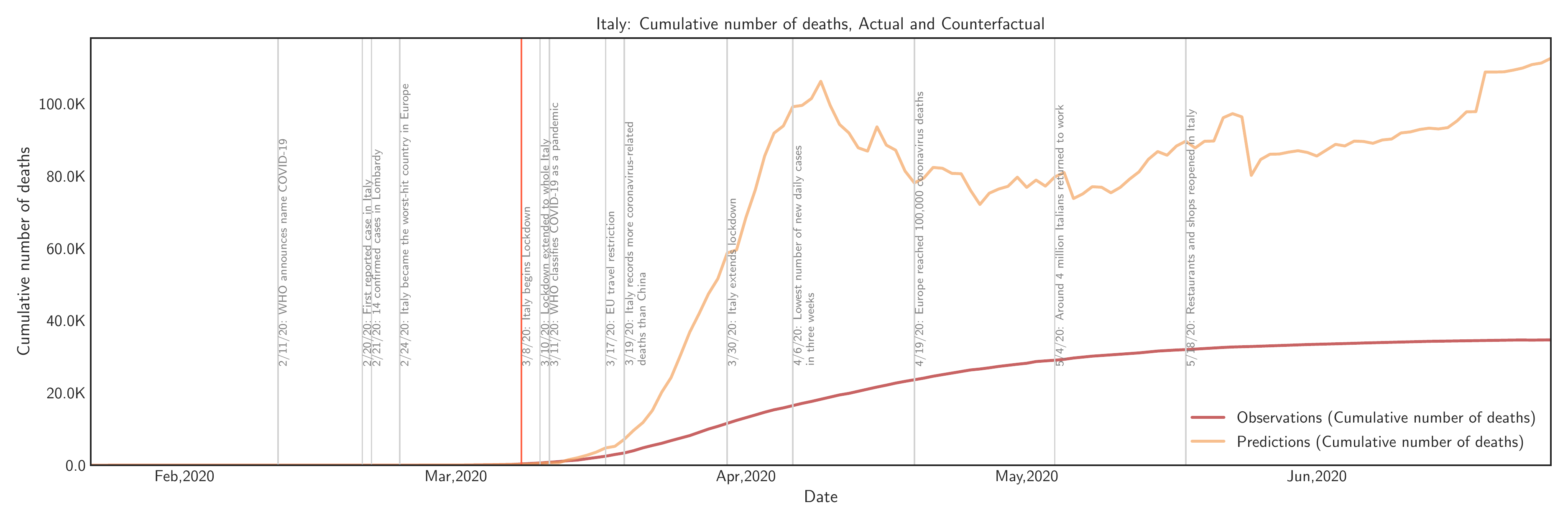}
	\end{subfigure}
	
	\begin{subfigure}[b]{\textwidth}
		\centering
		\includegraphics[width=0.9\linewidth,height=0.2\textheight]{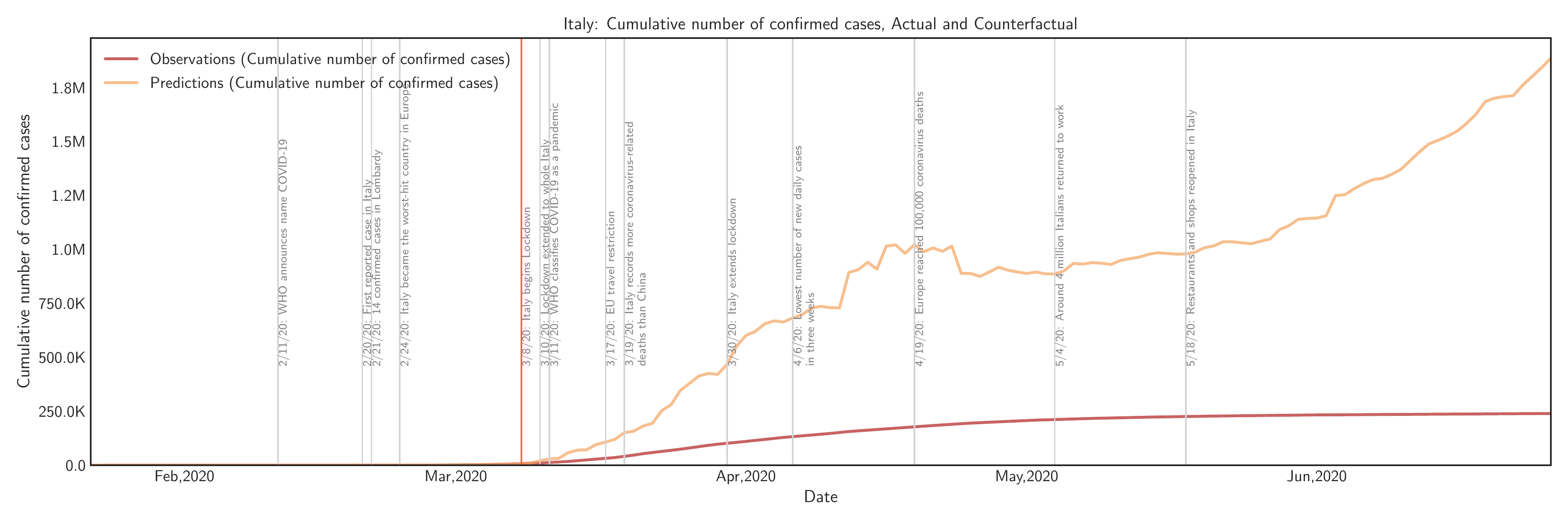}
	\end{subfigure}
	
	\caption[Italy]{Italy: (a) Actual number of deaths with linear trend line and points where trend changed (b) Actual number of deaths with counterfactual prediction estimated by m-RSC with actual lock-down date as intervention point (c) Actual number of confirmed cases with counterfactual prediction estimated by m-RSC.}
	\label{fig6} 
\end{figure*}

\subsection*{Delhi, India}
India had one of the most strict stay home order across the country in first phase of the lock-down between 25 March 2020 – 14 April 2020 (21 days), where an entire population of 1.3 billion people was put under restricted movement. Overall the lock-down had multiple phases, second phase was from 15th of April 2020 to 3rd of May 2020, and third phase was 4th of May to 17th of May, 2020. We present the counterfactual for each of these dates. However, in this case we consider both the daily number of confirmed cases as well as the cumulative number of confirmed cases for phase three of the lock-down. We limit the donor group as all others states of India.

\begin{figure*}
	\centering
	\begin{subfigure}[b]{\textwidth}
		\centering
		\includegraphics[width=0.8\linewidth,height=0.18\textheight]{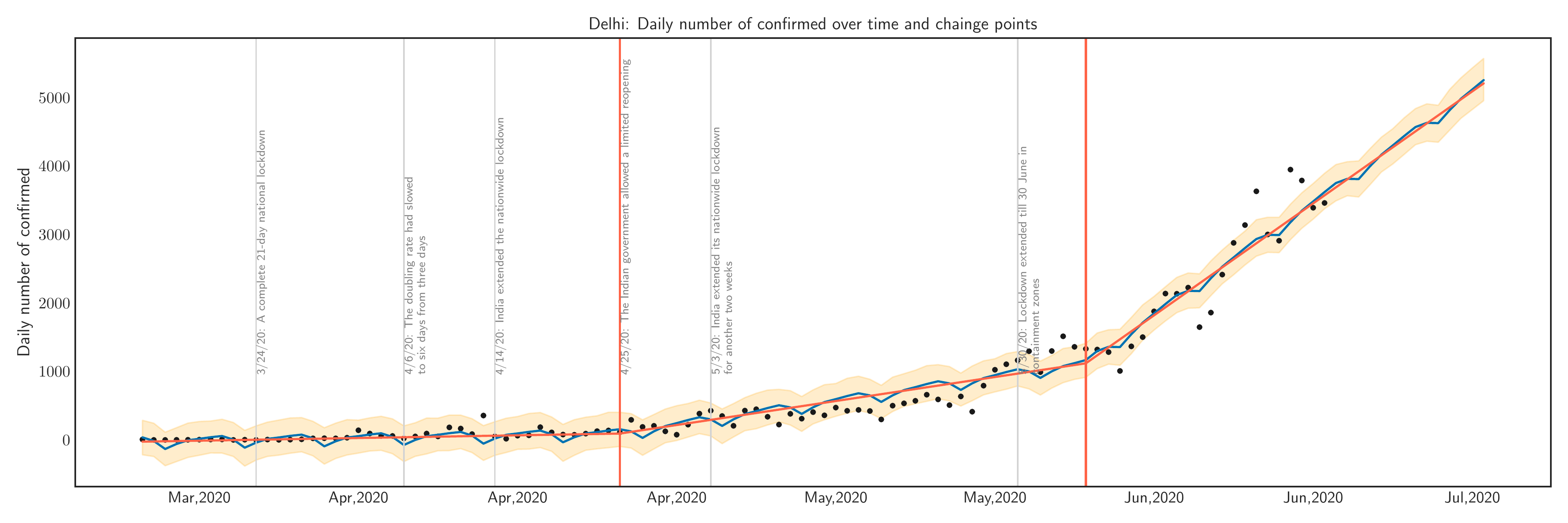}
	\end{subfigure}
	
	\begin{subfigure}[b]{\textwidth}
		\centering
		\includegraphics[width=0.8\linewidth,height=0.18\textheight]{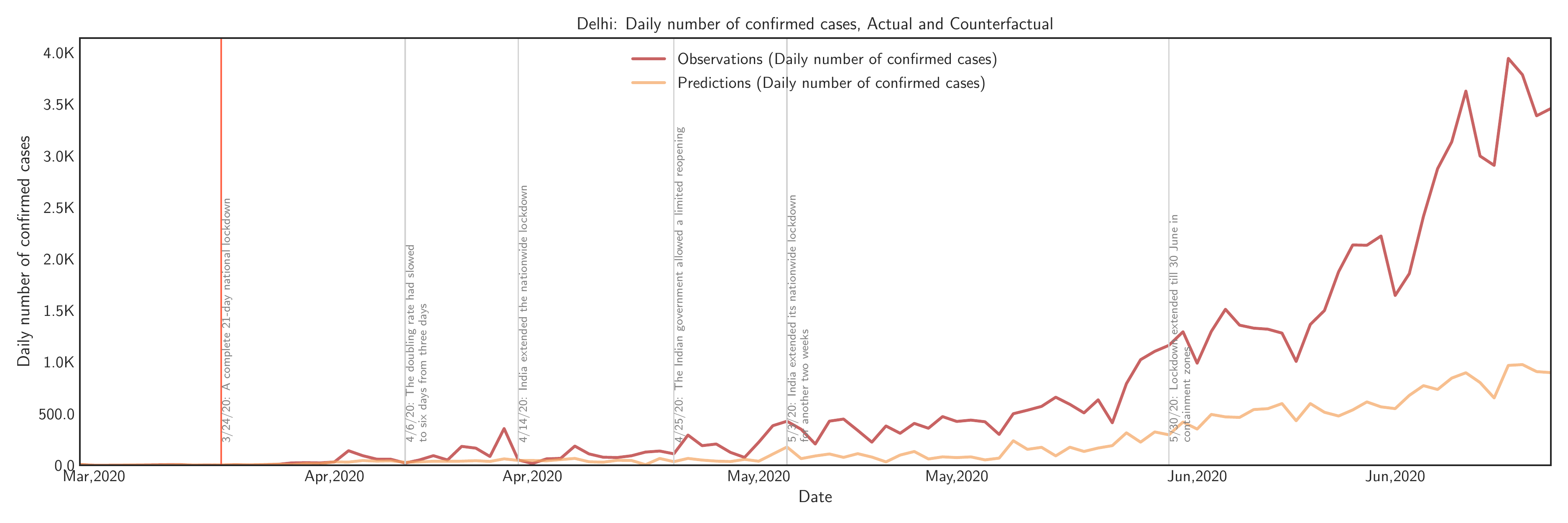}
	\end{subfigure}
	
	\begin{subfigure}[b]{\textwidth}
		\centering
		\includegraphics[width=0.8\linewidth,height=0.18\textheight]{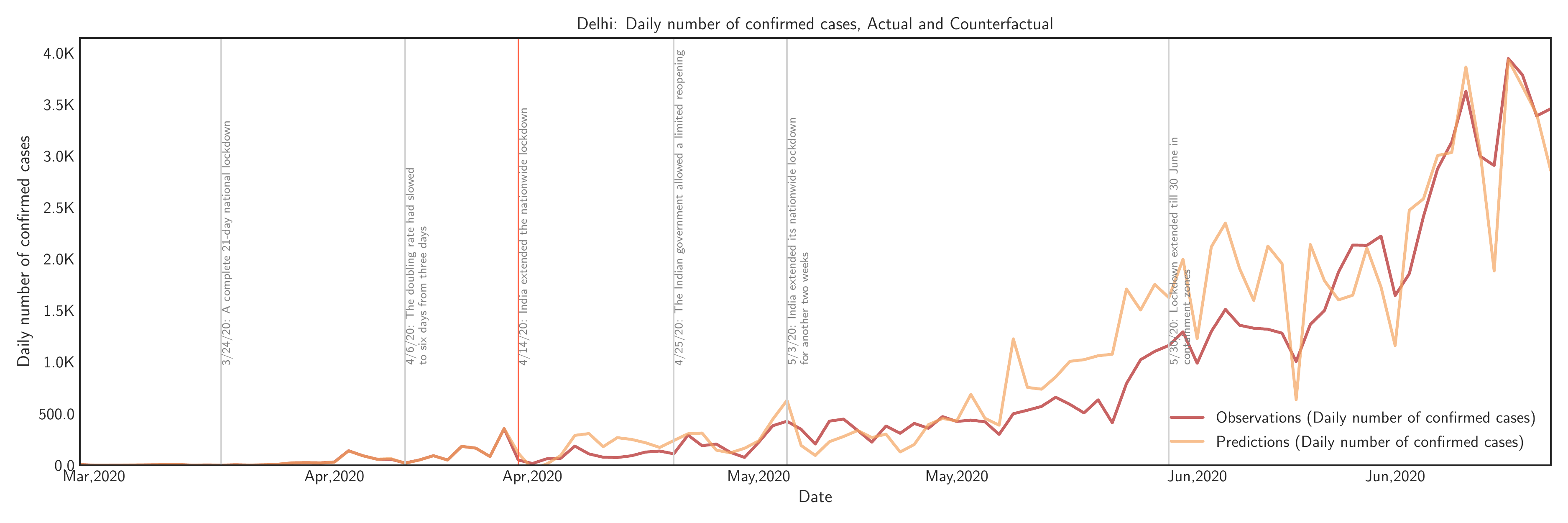}
	\end{subfigure}
	
	\begin{subfigure}[b]{\textwidth}
		\centering
		\includegraphics[width=0.8\linewidth,height=0.18\textheight]{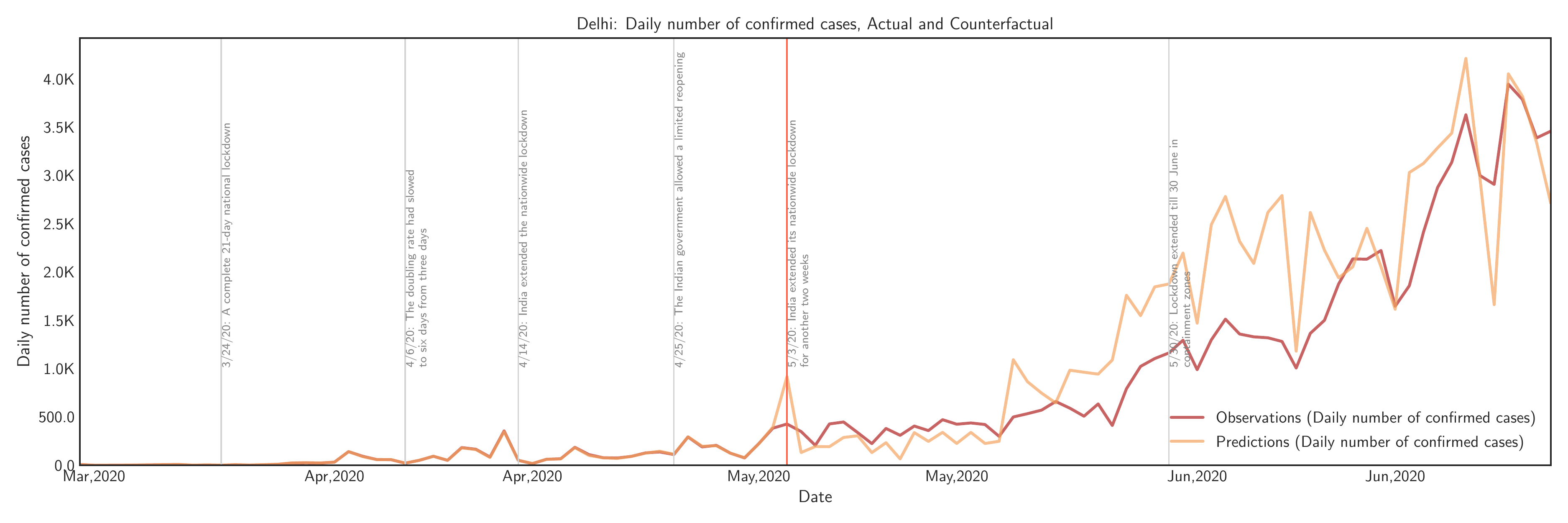}
	\end{subfigure}
	
	\begin{subfigure}[b]{\textwidth}
		\centering
		\includegraphics[width=0.8\linewidth,height=0.18\textheight]{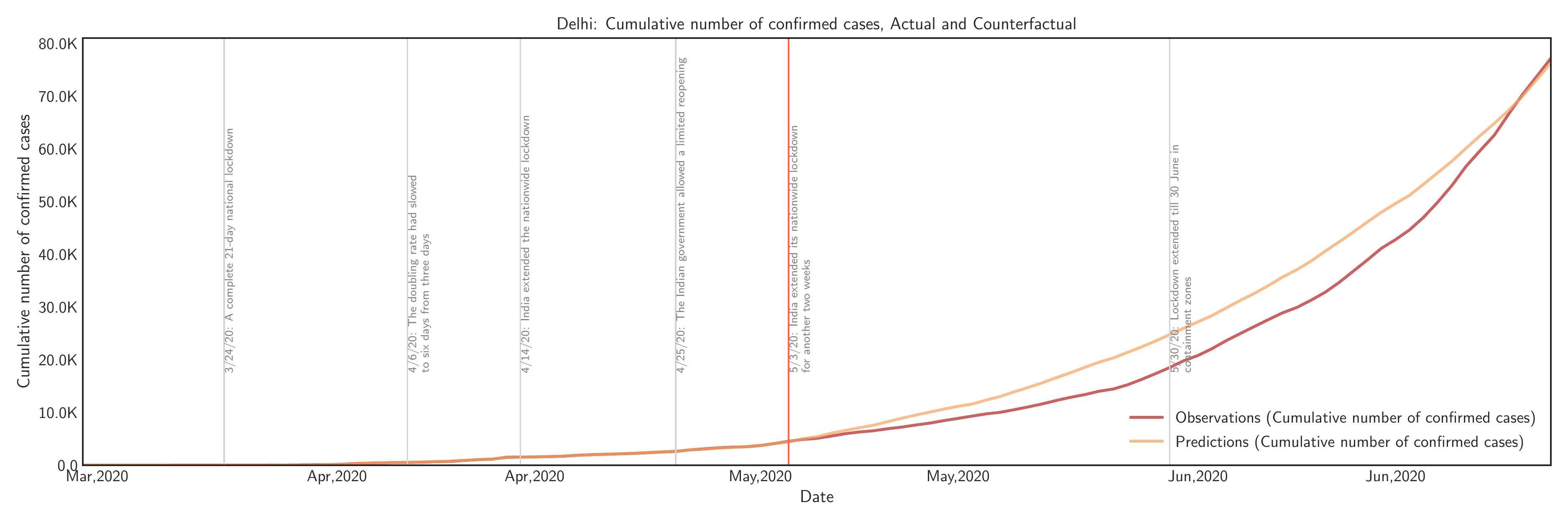}
	\end{subfigure}
	
	\caption[Delhi]{Delhi, India: (a) Actual number of confirmed cases with  trend line and points where trend changed (b) Actual number of confirmed cases with counterfactual prediction estimated by m-RSC with First lock-down date as intervention point (c) On Second lock-down date (d) On third  lock-down date}
	\label{fig7} 
\end{figure*}

Figure \ref{fig7} shows that counterfactual converges closely with the actual at third phase of the lock-down.  It should be noted that there are a few discrepancies in reporting. First, there is a weekly seasonality - possibly due to a lesser number of reports over the weekends. Second due to a revised higher number of reports on certain dates (high peak).

Linear trend-line fit shows two change points in the growth of number of cases - indicating that the exponential phase came at much later date and growth of the epidemic was under effective control in the earlier stages. Since, the stay-home order was applicable across all states and adherence was almost uniform -  trajectory of actual and counterfactual remains nearly same.

\section*{Discussion}
\label{SEC5}
In this work we use Multi-dimensional Robust Synthetic Control to understand the effects of stringency measure on COVID-19 pandemic. We construct synthetic version of a location using convex combination of other geographic locations in the donor pool that most closely resembled the treatment unit in terms of pre-intervention period using stringency index and adherence score (using mobility information). Results has been compared for The State of New York, Italy and Delhi, India with actual metric to that of counterfactual predicted by the algorithm.

In order to assess the robustness of the predictor we have computed MAPE and MdAPE measures and have shown their convergence to less than 20\% absolute error rate. In the future we would like to include  additional predictors like testing data, and virus strain information as they become available. Another direction of this study is to include parametric epidemic models like SIR-F\cite{Siettos2013}, and compare with m-RSC.

\section*{Methods}
	\label{SEC2}
		\subsection*{Tools}
	As stated above, our objective is to study the effects of government response at an aggregate level in terms of lives saved, and limiting the number of cases that require hospitalization. Such interventions can effectively be studied at a comparative level. In other words, if we have data for the evolution of aggregate outcomes, e.g. the number of confirmed cases and deaths, when policy is applied in a group under study versus when the same policy is not applied in a control group. However, government policies were applied at different level across a geographic region.  We do not have a mechanism to conduct a randomized trial. Hence, we consider using the synthetic control method \cite{ap08746, JMLR18, AMSS19}. In a synthetic control set up, where observational data is available for different groups, we can construct a synthetic or virtual control group by combining measurements from alternatives (or donors). In the following, we provide a brief overview of m-RSC \cite{AMSS19}.\par
	
	Suppose that observations from $N$ different geographically distinct groups or units are indexed by $i \in [N]$ in $T$ time periods (days) indexed by $j \in [T]$. Let $k \in [K]$ be the metrics of interest (e.g. number of confirmed cases, number of deceased, number of tests conducted, etc.). By $M_{ijk}$ we denote the ground-truth measurement of interest, and by $X_{ijk}$, an observation of this measurement with some noise. Let $1 \leq T_0 \leq T$ be the time instance in which our group of interest experiences an intervention, namely a government response to control the spread (e.g. stay home order, school or business closure, or mass vaccination). Without loss of generality we consider unit $i = 1$ (say, New York) and metric $k = 1$ (say, number of deaths) as our unit and metric of interest respectively.\par
	
	Our objective now is to estimate the trajectory of metric of interest $k = 1$  for unit $i = 1$ if no government response to control the spread had occurred. In order to do that we will use the trajectory associated with the donor units ($2 \leq i \leq N$ ), and metrics $k \in [K ]$. In the following we make two assumptions: (1) for all $2 \leq i \leq N$, $k \in [K]$ and $j \in [T]$, we have  $X_{ijk} = M_{ijk} + \epsilon_{ijk}$ where, $\epsilon_{ijk}$ is the observational noise, and (2) Same model is obeyed by $i=1$ in pre--intervention period, i.e. for all $j \in [T_0]$ and $k \in [K]$ we have $X_{1jk} = M_{1jk} + \epsilon_{1jk}$. As described by authors in \cite{AMSS19}, in following we also assume that for unit $i=1$, we only observe the measurement $X_{1jk}$ for pre-intervention period, i.e. for all $j \in [T_0]$ and $k \in [K]$. Our objective is to compute a counterfactual sequence of observation $M_{1jk}$ for the time period $j \in [T]$, and $k \in [K]$, and in specific for $T_0 \leq j \leq T$, and $k = 1$, using synthetic version of unit $i=1$.\par
	
	Define $\mathcal{M} = [M_{ijk}] \in \mathbb{R}^{N \times T \times K}$. $\mathcal{M}$ is assumed to have a few well behaved properties as required by the algorithm, namely, (1) $\mathcal{M} $ must be approximately low-rank and (2) every element $\left|M_{ijk}\right|$ shall have boundedness property (for details see \cite{AMSS19}). To check whether our model assumption holds in practice, we consider $N=185, T=150, K=2$, with $185$ countries as units. We consider number of confirmed cases and number of deceased as two metrics over $150$ days between January 22, 2020 and June 20, 2020. For assumption to hold, data matrix corresponding to number of confirmed cases and number of deceased and their combination should be approximated by a low-rank matrix. 
	
	\begin{figure*}
		\centering
		\includegraphics[width=0.8\textwidth]{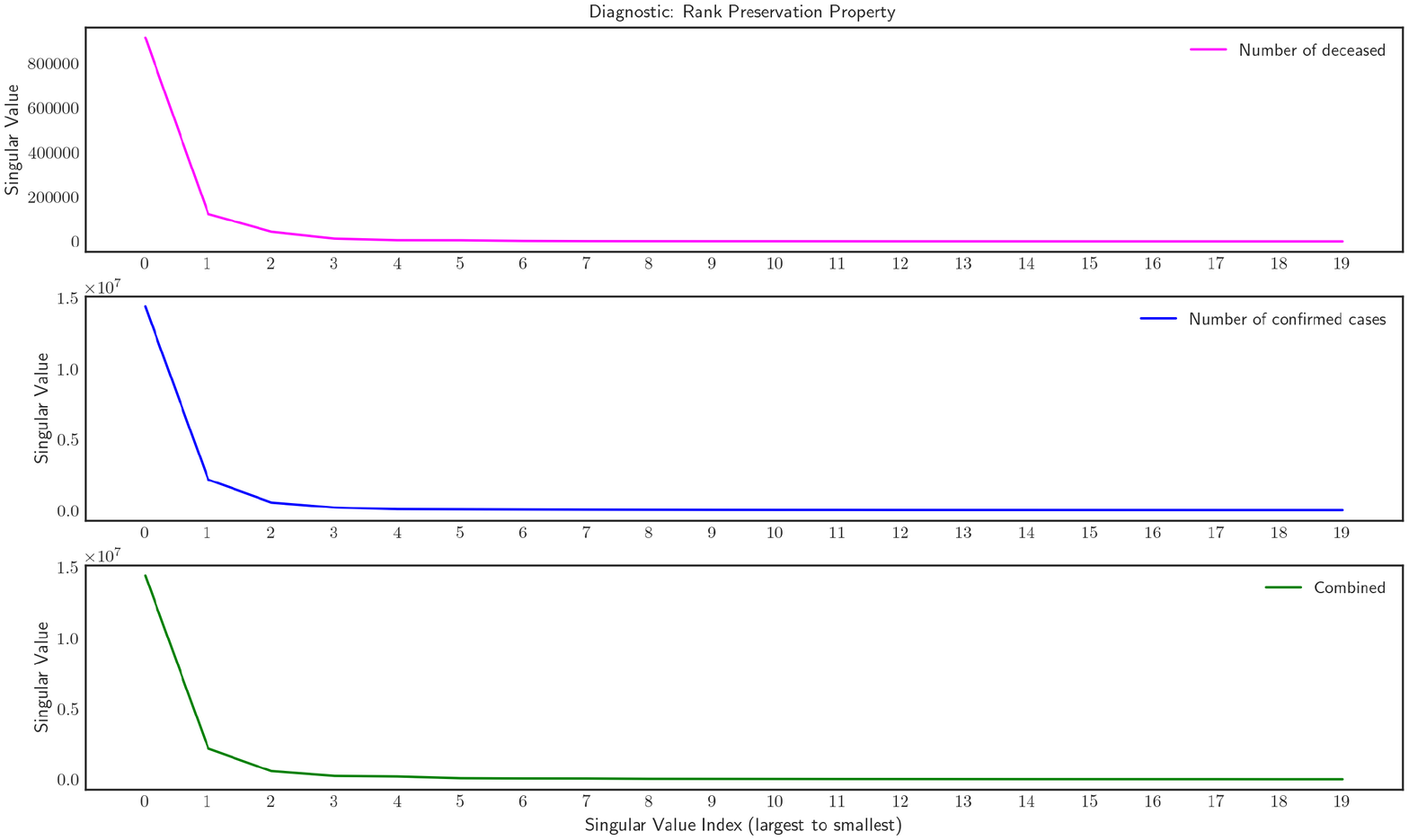}
		\caption{The Singular Value spectrum for all countries (of dimensions $185 \times 150$, showing top 20 singular values, in descending order.} 
		\label{fig1} 
	\end{figure*}
	
	As shown in Figure \ref{fig1}, the spectrum of the top 20 singular values (sorted in descending order) for each matrix. The plots clearly support the implications that most of the spectrum is concentrated within the top 5 principal components. The same conclusion holds true when units are states of The United States, and when we consider only countries in The  European Union.\par
	
	Let $\mathcal{Z} \in \mathbb{R}^{(N-1) \times T \times K}$ corresponding to donor units, and $X_1 \in \mathbb{R}^{1 \times T_0 \times K}$ correspond to unit under intervention. We obtain $\hat{\mathcal{M}}$ from $\mathcal{Z}$ after applying a hard singular value thresholding. Subsequently, weights are learned using linear regression by computing
	
	\begin{equation*}
	\hat{\beta} = \argmin_{v \in \mathbb{R}^{(N-1)} } \left\| X_1 - v^T \hat{\mathcal{M}}_{T_0}\right\|^2_2
	\end{equation*}
	
	For every $k \in [K]$, the corresponding estimated counterfactual means for the treatment unit is then defined as
	
	\begin{equation*}
	\hat{\mathcal{M}}_1^{(k)} = \hat{\beta}^T \hat{\mathcal{M}}^{(k)}
	\end{equation*}
	
	\subsection*{Measures of stringency and mobility}
	As described in Section \ref{SEC1}, we use m-RSC to construct a synthetic control for the treatment unit using data from multiple control units or donor group using pre-intervention period data.  The synthetic control is then used for estimating the counterfactual in the post-intervention period. In our setup, intervention date is typically the date when a stay-home order or lock-down was declared for the treatment unit.  However, government policy may have been applied over time with different levels of stringency measures. 
	
\begin{figure*}
	\centering
	\begin{subfigure}[b]{\textwidth}
		\centering
		\includegraphics[width=0.7\linewidth,height=0.15\textheight]{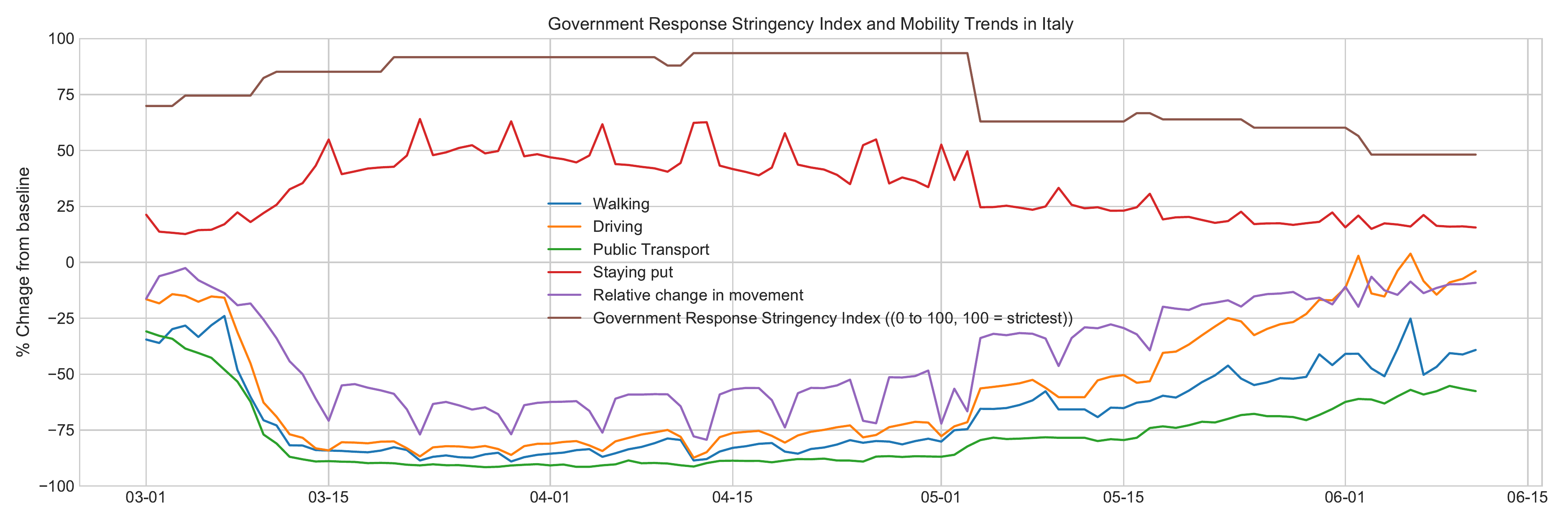}
	\end{subfigure}
	
	\begin{subfigure}[b]{\textwidth}
		\centering
		\includegraphics[width=0.7\linewidth,height=0.15\textheight]{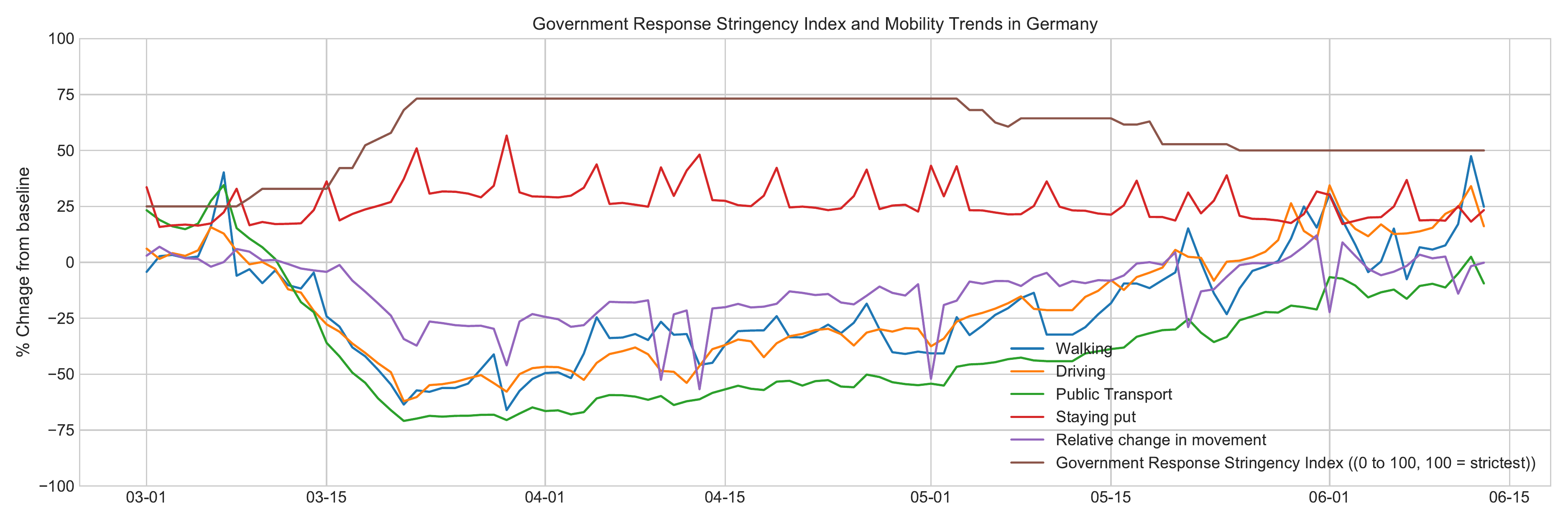}
	\end{subfigure}
	
	\begin{subfigure}[b]{\textwidth}
		\centering
		\includegraphics[width=0.7\linewidth,height=0.15\textheight]{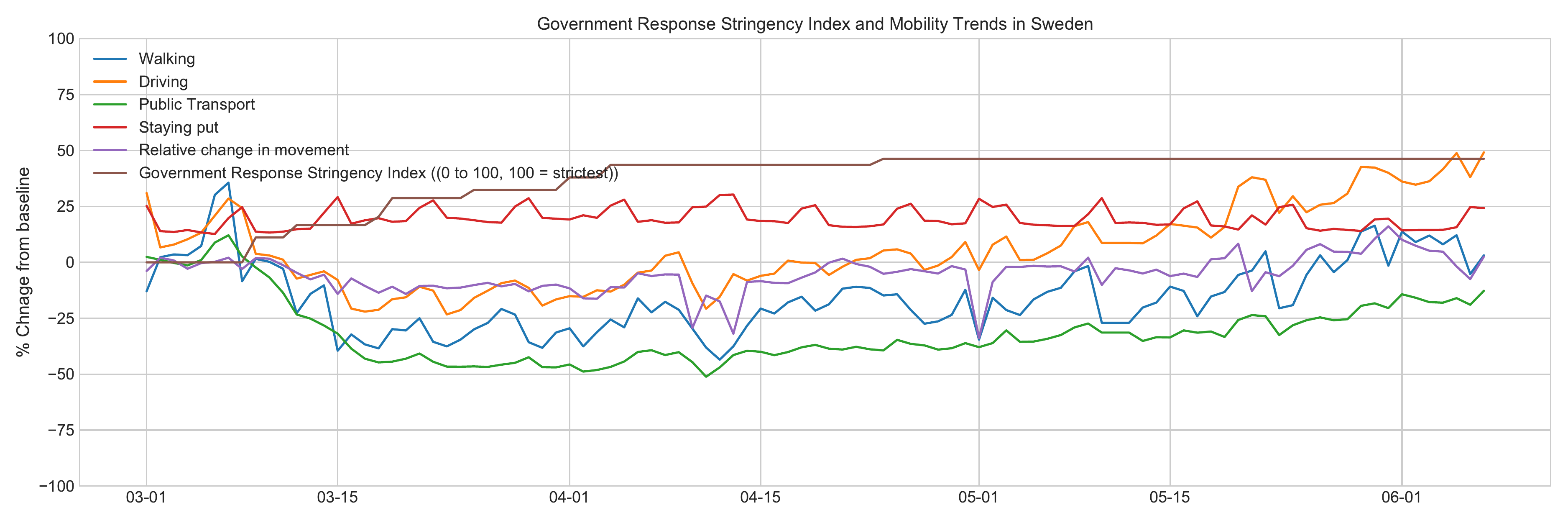}
	\end{subfigure}

	\caption[Mobility Trends]{Stringency index, and Mobility Trends in (a) Italy (b) Germany (c) Sweden}
	\label{fig2} 
\end{figure*}

	To understand this we use stringency and policy indices data from OxCGRT \cite{HWP2020}, which records the strictness of policies that restrict people’s behavior and includes 8 different measures - e.g. school  and workplace closure, cancellation of public events, restrictions on gathering size etc. Figure \ref{fig2}, shows the plot of stringency index, with mobility data.

    It can be observed that the level of lock-down varies over time and geographic region. We use this information in two different ways. First we choose the maximum level of index, first increased level of restriction index, and 14 days after the maximum level of index as various intervention dates and compare their effects. Second, we combine this information with mobility data to select the control groups for a treatment unit as discussed below.
    
    In order to understand the effect of stringency on a treatment unit for a metric, we need to select a donor group where level of stringency was different or adherence to stringency was different. Since, there was a degree of stringency and adherence to such measures at different level - under any possible choice of donor group, we acknowledge, that we will be underestimating the counterfactual - i.e. what would have been without any stringency measures. To estimate the degree of adherence to lock-down measures, we use mobility data from Apple, Google and Facebook. Apple mobility data provides a relative volume of directions requests per region, sub-region or city compared to a baseline volume - i.e. percentage change over time from the baseline including weekly seasonality. Facebook data provides the relative percentage of population that is staying in the same place and also the percentage of population that moved from a region to another.  In Facebook data, to quantify how much people move around measure is derived by counting the number of level-16 Bing tiles (which are approximately 600 meters by 600 meters in area at the equator) they are seen in within a day. Assuming $U_{d,r}$ is the set of eligible users in region $r$ on day $d$, and $tiles(u)$ is the number of tiles visited by a given user $u$ in $U_{d,r}$, total number of tiles visited for that region is given by $totaltiles(U_{d,r}) = \sum_{u \in U_{d,r}} min(tiles(u), 200)$. Change in Movement measure is then the difference between a baseline and value on day $d$ for average number of $totaltiles(U_{d,r})$. Similarly, Stay Put metric is calculated as the percentage of eligible people who are only observed in a single level-16 Bing tile during the course of a day on an average compared to a baseline. Finally, Google Community Mobility Report provides a percent change in visits to places like grocery stores and parks within a geographic area from baseline.
    
    It can be seen from Figure \ref{fig2}, that the adherence  and stringency level do not correspond. For example, in Sweden, with increasing level of government measures between March - April, there has not been any significant change in proportion of people staying put or moving between regions. Similarly, in other places, it can be observed that while government measures remain at the same level over April, number of people staying put at one place starts declining. 
    
\begin{figure*}
    	\centering
    	\begin{subfigure}[b]{\textwidth}
    		\centering
    		\includegraphics[width=0.7\linewidth]{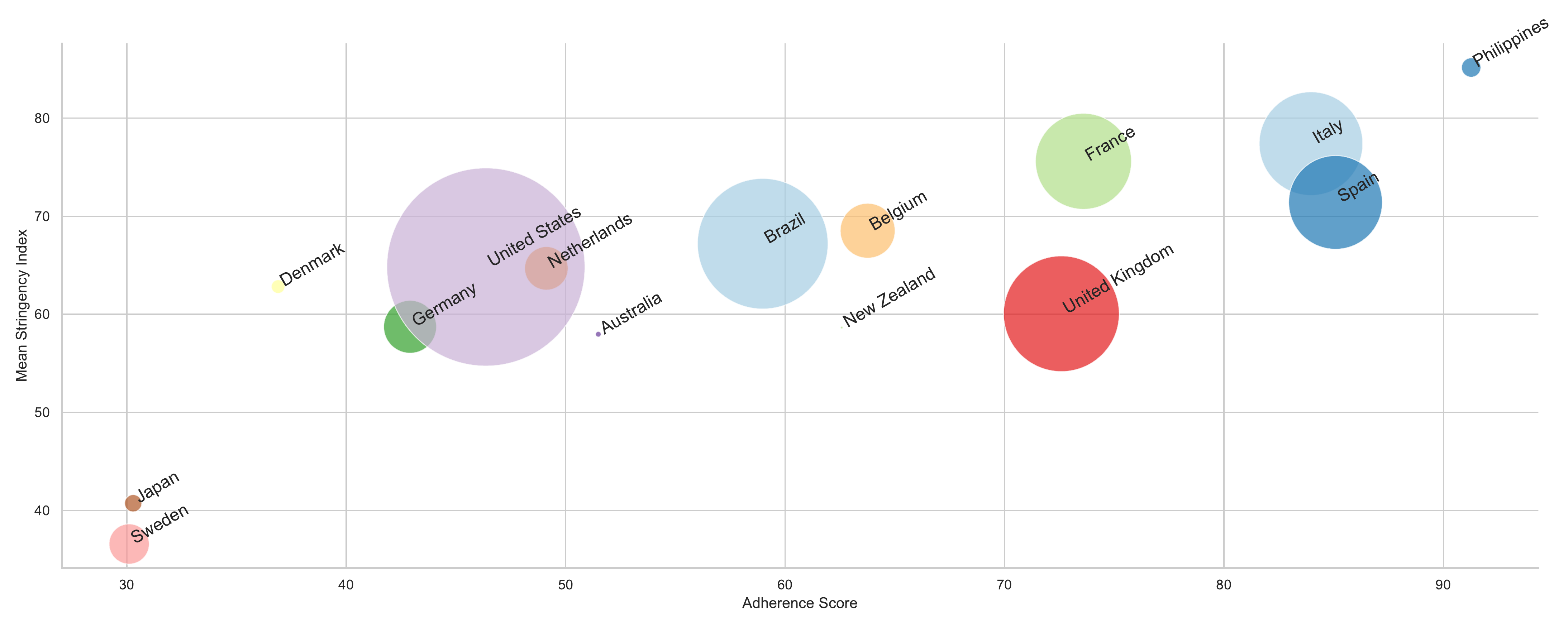}
    	\end{subfigure}
    	
    	\begin{subfigure}[b]{\textwidth}
    		\centering
    		\includegraphics[width=0.7\linewidth]{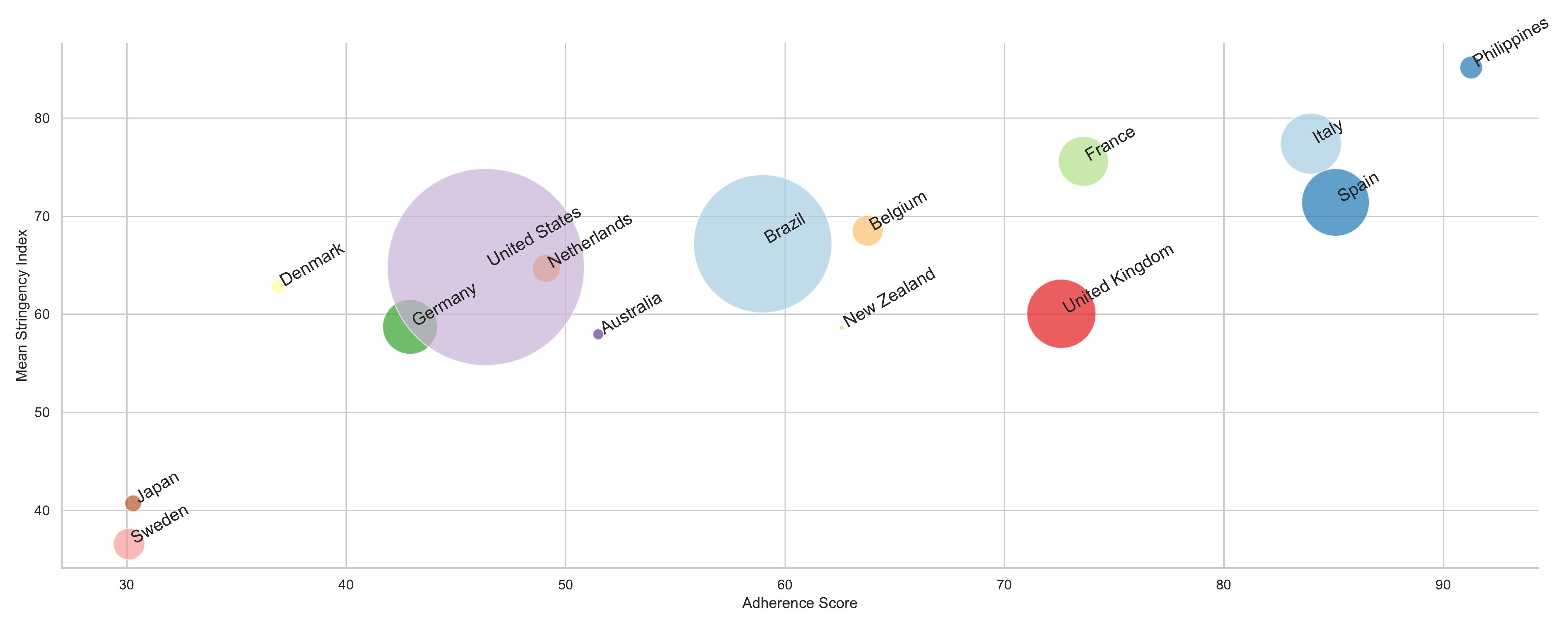}
    	\end{subfigure}
    	
    	\caption[Mobility Trends]{Adherence score vs. Stringency index where size of each circle is determined by (a) total number of deaths till date (b) total number of confirmed cases till date.}
    	\label{fig3} 
    \end{figure*}

    Selecting donor group: We combine metrics that allow the spread of the virus and similarly combine those that reduces the possibility of spread. We combine them by taking average define a single adherence score. For unit $i \in [N]$, $\forall j \neq i : j \in [N]$, $j$ is a donor unit if adherence score, and stringency Index of $j$ is less than $i$. Figure \ref{fig3} shows this relation in graphical form for a selected set of countries. As per Figure \ref{fig3}, United States, Brazil and United Kingdom are in the donor group when we compute counterfactual estimate for Italy. 
    
\begin{figure*}
		\centering
		\includegraphics[width=0.7\textwidth]{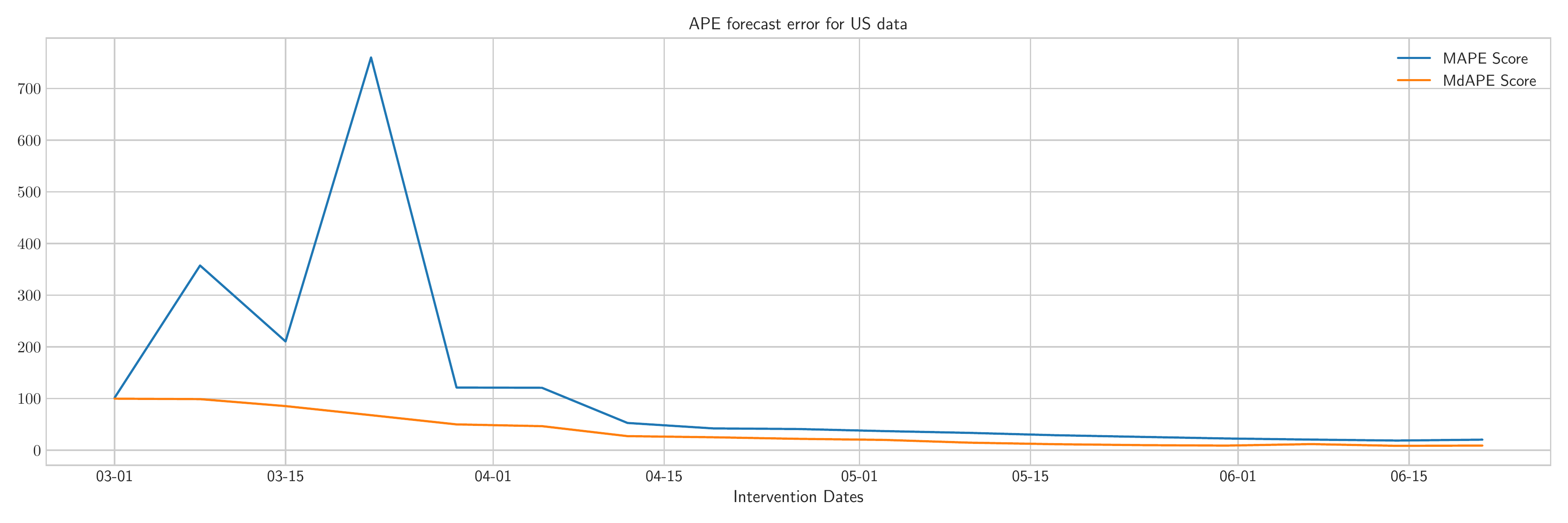}
		\caption{MAPE} 
		\label{fig4} 
	\end{figure*}

	Statistical Performance Evaluation: in Figure \ref{fig4} we present the distribution of Mean and Median absolute percentage error statistic for the runs-forecasts from m-RSC algorithm with changing forecast horizon. We consider every Mondays between March 1, 2020 to June 21, 2020, both included to forecast the number of deaths at the end of the day on June 26, 2020 for all states in United States, and compare with actual data. For every state, donor group is selected using the method described above. 
    It can be seen from Figure \ref{fig4}, that both Mean and Median absolute percentage error statistic are larger, when larger forecast horizon has been considered, and that is to be expected. Both MAPE and MdAPE converge to a less than 20 percent when horizon is about a week. MAPE being very large from MdAPE clearly indicates a right skew in the predicted values. This exactly corresponds to the places where exponential growth of the pandemic can be observed by March 22, 2020. The counterfactual prediction for June 26, 2020 on March 22, 2020 for these few places are several times higher as as expected, while, stricter stringency measures were being implemented in The United States around those dates.
    
	\subsection*{Data Source}
	We use following five sources of data as described in Table \ref{Table1}:
	
	\begin{table*}
		\centering
		\begin{tabularx}{0.9\textwidth}[t]{p{0.3\textwidth}X}
			\hline
			\textbf{Data source} &  \textbf{Description}\\ [0.5ex]
			\hline\hline
			Daily update from JHU  \cite{DDG2020}  & We use this data to derive metric for units: i.e. number of confirmed cases and number of deceased for each day and geographic locations.\\ [1ex]
			\hline
			Facebook Movement Range Maps \cite{JM2020} & The relative percentage of population that is staying put and also the percentage of population that moved from a region to another. We use this data in selection of donor units.\\
			\hline
			Apple Mobility Trends\cite{Apple2020} & Relative volume of directions requests per region, sub-region or city compared to a baseline volume, categorized by Driving, Walking or Public Transport. We use this data in selection of donor units.\\
			\hline
			OxCGRT  \cite{HWP2020} &  Strictness of policies that restrict people’s behavior, 8 measures combined to provide a score between 0 and 100, where 100 being most stringent. We use this data in selection of donor units.\\
			\hline
			Google Community Mobility Report \cite{Goog2020} &  Percent change in visits to places like grocery stores and parks within a geographic area. We use this data in selection of donor units.\\ [1ex] 
			\hline
		\end{tabularx}
		\caption{Data Source}
		\label{Table1}
	\end{table*}
	 
	\section*{Authors' contributions}
	
	S. K. G.: Conceptualization, Data Collection, Software, Visualization, Writing - Original Draft, Writing - Review and Editing. S. G.: Writing - Original Draft, Writing - Review and Editing. S. S. N.: Data collection, Writing - Review and Editing. All authors reviewed the manuscript.
	
	\section*{Data and Code}
	All data and code used for this work is made available here: \url{https://github.com/subhaskghosh/lockdown-paper}
	
	\section*{Competing Interests statement}
	The authors have no competing interests.
	
	\bibliography{lockdown.bib}	
\end{document}